\documentclass[english]{article}
\usepackage[T1]{fontenc}
\usepackage[latin1]{inputenc}
\usepackage{a4}
\usepackage{amsmath}
\usepackage{babel}
\usepackage{graphics}
\usepackage{setspace}
\usepackage{amssymb}

\makeatletter

\providecommand{\LyX}{L\kern-.1667em\lower.25em\hbox{Y}\kern-.125emX\@}
\let\SF@@footnote\footnote
\def\footnote{\ifx\protect\@typeset@protect
    \expandafter\SF@@footnote
  \else
    \expandafter\SF@gobble@opt
  \fi
}
\expandafter\def\csname SF@gobble@opt \endcsname{\@ifnextchar[%]
  \SF@gobble@twobracket
  \@gobble
}
\edef\SF@gobble@opt{\noexpand\protect
  \expandafter\noexpand\csname SF@gobble@opt \endcsname}
\def\SF@gobble@twobracket[#1]#2{}

\usepackage[T1]{fontenc}
\usepackage[latin1]{inputenc}
\usepackage{babel}
\usepackage{graphics}


\begin{document}

\title{Topologically coupled energy bands in molecules}

\author{Frédéric Faure%
\thanks{Laboratoire de Physique et Modélisation des Milieux Condensés (LPM2C)
(Maison des Magisteres Jean Perrin, CNRS) BP 166 38042 Grenoble Cedex
9 France. \protect \\
email: frederic.faure@ujf-grenoble.fr \protect \\
http://lpm2c.polycnrs-gre.fr/faure/
}\\
Boris Zhilinskii%
\thanks{Université du Littoral, MREID, 145 av. M. Schumann, 59140 Dunkerque , France
zhilin@univ-littoral.fr
}}

\maketitle
\begin{abstract}
We propose a concrete application of the Atiyah-Singer index formula
in molecular physics, giving the exact number of levels in energy
bands, in terms of vector bundles topology. The formation of topologically
coupled bands is demonstrated. This phenomenon is expected to be present
in many quantum systems. 
\end{abstract}


\section{ Introduction}

Molecular spectra of small molecules possess thousands of energy levels
which are known with high precision and characterize the dynamical
behavior of nuclei and electrons in very fine details. Thus a qualitative
approach may be useful in front of the huge complexity of the spectrum
\cite{boris2,PhysRep2,Harter}. Due to symmetries of different origins,
the energy levels are often regrouped together in multiplets with
quasi-degeneracies, forming so called energy bands. In terms of dynamics,
the fine structure of the bands reflects a slow motion, whereas the
coarser repartition of different energy bands is related to a faster
motion. These two motions can be strongly coupled together. Several
typical molecular examples can be suggested to illustrate the presence
of such a band structure and the evolution of this band structure
with the variation of external parameters.

i) Vibrational structure of different electronic states. The existence
of a fine vibrational structure associated with different electronic
states is the standard consequence of the Born-Oppenheimer (adiabatic)
separation of electronic and nuclear variables in molecules based
on the smallness of the ratio \( (m_{e}/M_{n})^{1/4}\sim 1/10 \)
(here \( m_{e} \) and \( M_{n} \) are the electron mass and the
typical mass of nuclei) \cite{Simon}.

ii) Rotational structure of different vibrational states. The rotational
excitations are typically smaller than the vibrational ones. The quantum
spectrum has a fine rotational band structure associated with a more
crude vibrational structure. The absolute value \( J \) of the angular
momentum is a strict integral of motion for an isolated molecule,
and can be used as a natural parameter to analyze the evolution of
rotational bands \cite{PhysRep2,Harter}.

iii) A more sophisticated example considered in this paper arises
for systems which possess vibration modes with (quasi)-degenerate
frequencies. Such a situation typically takes place in problems with
symmetry. The description of the degenerate vibrations in the harmonic
oscillator approximation gives an energy spectrum formed by degenerated
energy levels, the so called polyads. Each polyad is characterized
by the total number of vibrational quanta \( N \), it contains. In
systems where these vibrational modes are coupled with electronic
states, one obtains an electronic-vibrational (vibronic) band structure.
A typical question is the evolution of this vibronic band structure
with \( N \) \cite{Kellman,GenF} and its persistence (or reorganization)
as a function of external physical parameters.

In all these previous examples, when an external parameter is varied,
along with modifications of the internal structure of each isolated
band, one observes more serious modifications of the global band structures
\cite{boris,key-1,key-2,borisA,borisB}. Sometimes two bands come
in contact. During this contact, a certain number of levels pass from
one band to the other, resulting in a redistribution of levels between
bands. These qualitative phenomena reflect band structure properties
which are robust under modification of parameters, except for abrupt
structure changes, and suggest a topological description. One expects
a characterization of each band which would give the precise number
of levels it contains, and would predict the possible band change
of structure under an external parameter variation. Such a characterization
has been done for rotational bands in \cite{fred-boris,key-3}. It
has been shown that an integer topological Chern index can be attributed
to each individual rotational band and a generic modification of the
band structures is related to the exchange of one energy level between
two consecutive bands. The presence of symmetry leads to more complicated
redistribution of energy levels between bands. The purpose of the
present paper is to describe a new qualitative phenomenon related
with the formation of topologically coupled bands.

The mathematical approach which we need to characterize these qualitative
modifications is a semi-classical analysis within the Born-Oppenheimer
approximation and topological analysis of vector bundles. Precise
mathematical theories are developed in \cite{hawkins00,fedosov2000,emmrich}.

 For rotational bands, the topological phenomena are quite simple
due to the fact that the slow rotational motion has one degree of
freedom, and takes place on a two dimensional sphere \cite{fred-boris,key-3}.
We present now a more interesting example both from the physical and
mathematical point of view, where the slow motion comes from molecular
vibrations and has two degrees of freedom (a four dimensional phase
space).

\section{Semi-quantum model with \protect\protect\( CP^{2}\protect \protect \)
classical phase space}

 We first consider three vibrations with the same frequency (1:1:1
resonance) described by\begin{equation}
\label{e:Hvib}
H_{vib}=\sum _{k=1,2,3}\frac{1}{2}\left( p_{k}^{2}+q_{k}^{2}\right) =\sum _{k=1,2,3}\left| Z_{k}\right| ^{2},
\end{equation}
 where a complex notation \( Z_{k}=\left( q_{k}+ip_{k}\right) /\sqrt{2} \)
is used. The time evolution of a point \( Z=\left( Z_{1},Z_{2},Z_{3}\right)  \)
in phase space \( {\mathbb {C}}^{3} \) is given by a phase factor:
\( Z(t)=\exp \left( -it\right) Z(0) \). The family of (closed) trajectories
of a given energy, say \( E=1 \), forms a four dimensional phase
space {[}\( 4=6-1 \)(energy) \( -1 \)(time){]} which is a reduced
phase space from the classical mechanics point of view. This space
is known in mathematics as the complex projective space and noted
\( {\mathbb {C}}P^{2}=P({\mathbb {C}}^{3}) \): a point in \( {\mathbb {C}}P^{2} \)
corresponds to a given closed trajectory of energy \( E \) and will
be noted by \( \left[ Z\right]  \) in the sequel. The corresponding
quantum Hamiltonian \( \hat{H}_{vib}=\sum _{k=1,2,3}a^{+}_{k}a_{k}+3/2 \)
is obtained by replacing \( Z_{k},\overline{Z_{k}} \) by operators
\( a_{k},a_{k}^{+} \) respectively. A given energy level is thus
an integer \( N \) plus \( 3/2 \), and this level is degenerated
with multiplicity \[
{\mathcal{N}}_{0}=\left( N+1\right) \left( N+2\right) /2.\]
 The associated eigenspace \( {\mathcal{H}}_{polyad}={\mathbb {C}}^{{\mathcal{N}}_{0}} \)
is called the polyad \( N \), and corresponds to the classical phase
space \( {\mathbb {C}}P^{2} \) \cite{boris89}.

 Let us now consider a more complicated system where the previous
degenerated vibrations are coupled with additional degrees of freedom.
An explicit model will be given in the next section. We first discuss
physical situations in molecular physics which motivate this model.

 Suppose that these additional degrees of freedom have much higher
energy excitations, which can be related, for example, to electronic
excitation, or to another vibrational excitation with higher frequency.
Due to this higher scale of energy, it is sufficient to take into
account only finite (and rather low in fact) number of quantum states
of this second subsystem, called the {}``fast'' subsystem. In the
absence of quasi-degeneracies in the fast subsystem one state should
be sufficient, but in the case of the presence of degeneracies or
quasi-degeneracies we should naturally treat together all the quasi-degenerate
quantum levels.

Such a situation takes place, for example, for the dynamic Jahn-Teller
effect of type \( F-f \) \cite{englman}, which describes the coupling
of the triply degenerate vibrational modes, with a multiplet of triply
degenerate electronic states. Although in the standard treatment of
Jahn-Teller effect the coupling between vibrational states from different
polyads is explicitly taken into account \cite{englman,Hougen,Judd},
in the limit of weak Jahn-Teller coupling, the restriction to the
effective model for an isolated polyad is quite natural \cite{Boudon,key-4}.
We deal with this effective model here.

Another similar and natural example is the vibrational structure of
combination bands \( \nu _{3}+N\nu _{4} \) in tetrahedral AB\( _{4} \)
type molecules. Due to tetrahedral symmetry both vibrational modes
\( \nu _{3} \) and \( \nu _{4} \) are triply degenerate. The \( \nu _{3} \)
frequency which is mainly due to A-B stretching vibrations is typically
higher than the \( \nu _{4} \) frequency, associated with bending.
In the absence of resonance between \( \nu _{3} \) and \( \nu _{4} \)
we can treat the combination band \( \nu _{3}+N\nu _{4} \) as formed
by three bands associated with three quantum states corresponding
to one quantum excitation of triply degenerate \( \nu _{3} \) mode.
Each of these quantum states is associated with the sequence of vibrational
polyads formed by \( \nu _{4} \) overtones.

 In both cases, purely vibrational and Jahn-Teller, the mathematical
model can be formalized in the same way. We consider the coupling
of three quantum states with vibrational polyads formed by triply
(quasi)degenerate modes. We associate the three quantum states with
the high-frequency motion (it is the {}``fast'' subsystem) and vibrational
polyads with low-frequency motion (it is the {}``slow'' subsystem).
To simplify the language we will speak about {}``fast'' subsystem
as electronic and about {}``slow'' subsystem as vibrational. The
general structure of this model can be described as {}``semi-quantum''
model. The notion {}``semi-quantum'' used in some of earlier publications
\cite{borisA,borisB} reflects the presence of two subsystems, one
is treated as classical whereas another as purely quantum.

 The simplest model assumes the degeneracy of electronic states and
the degeneracy of vibrational states within polyads. Among the possible
physically interesting corrections there are:

 i) vibrational couplings and anharmonicities which result in splitting
of degeneracies in vibrational polyads, but make no distinction between
different electronic states;

ii) pure electronic splitting, which does not influence the internal
vibrational polyad structure and can be present either (in the absence
of symmetry) due to initial splitting of accidentally quasi-degenerate
electronic states or produced by symmetry breaking effects like external
electric or magnetic fields;

 iii) vibronic coupling between electronic and vibrational states.

 We will suppose that the purely vibrational coupling is not important,
because it modifies only the internal structure of energy bands. Two
really different limit cases are ii) and iii) above. They correspond
to situations with, on one hand, purely electronic splitting being
large as compared to vibronic coupling and on the other hand with
the vibronic coupling being the most essential. Naturally, this last
case corresponds to strictly degenerate (for example due to symmetry)
electronic states.

 The simultaneous presence of a pure electronic splitting and a vibronic
coupling will be described by an explicit model in the next section,
eq.(\ref{ParFam}). We will introduce an external parameter \( \lambda =0\rightarrow 1 \),
which will relate the two extreme situations ii) and iii).

\subsection{A matrix model}

 Let us describe three {}``electronic'' states by the quantum Hilbert
space \( {\mathcal{H}}_{elec}={\mathbb {C}}^{3} \). We suppose that
vibronic (electronic-vibrational) coupling is sufficiently small as
compared to vibrational frequency and can be taken into account only
through interactions between polyads with the same polyad quantum
number \( N \) which belong to different electronic states. This
means that the total Hilbert space, we are interested in, \( {\mathcal{H}}_{tot}={\mathcal{H}}_{polyad}\otimes {\mathcal{H}}_{elec} \)
has dimension \( 3{\mathcal{N}}_{0} \).

 The main effect of the coupling is to remove the degeneracy of the
triplets of vibrational polyads. In the {}``semi-quantum'' description
of Born-Oppenheimer approximation, when the vibrational motion is
treated classically, and the electronic motion is still quantum, the
coupling results in a slow precession of the orbits, that is a slow
motion of the representative point in \( {\mathbb {C}}P^{2} \). We
will introduce the simplest form of the coupling, in a way similar
to the standard spin-orbit coupling \( {\mathbf{S}}\cdot {\mathbf{L}} \),
but now the coupling is chosen to be invariant with respect to the
SU(3) symmetry group. For that purpose, we extend the SU(3) symmetry
of the three-dimensional harmonic oscillator (\ref{e:Hvib}) to the
three electronic states, supposing that the space \( {\mathcal{H}}_{elec} \)
is the three-dimensional fundamental representation of SU(3), and
that the Hamiltonian \( \hat{H}_{1} \) below is SU(3) invariant.

 In the spirit of semi-quantum approach, a classical Hamilton function
is defined on phase space \( {\mathbb {C}}P^{2} \) with values in
the space of \( 3\times 3 \) Hermitian matrices:\[
\left( H_{1}\left( \left[ Z\right] \right) \right) _{i,j=1,2,3}=\overline{Z_{i}}Z_{j}/\left( \sum _{k=1,2,3}\left| Z_{k}\right| ^{2}\right) =\left( \left| Z\left\rangle \right\langle Z\right| \right) _{i,j}.\]
 The last expression shows that \( H_{1}\left( \left[ Z\right] \right)  \)
can be interpreted geometrically as a projector in \( {\mathcal{H}}_{elec} \)
space onto the one dimensional space, a line, defined by the value
of \( \left[ Z\right]  \). From this, the eigenvalues of \( H_{1}\left( \left[ Z\right] \right)  \)
are \( E_{Line}=1 \), and \( E_{Orth}=0 \) with multiplicity two.
{}``\( Orth \)'' is for the two dimensional space orthogonal to
the line.

\begin{figure}[htbp]
{\centering \resizebox*{1\columnwidth}{!}{\includegraphics{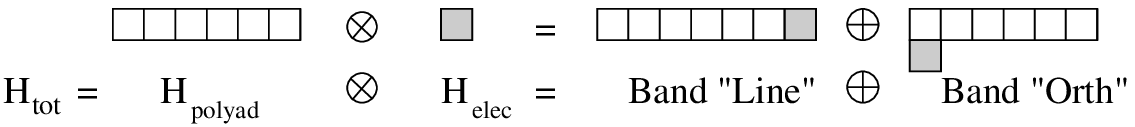}} \par}

\caption{\label{f:fig1}Spectrum of \protect\( \hat{H}_{1}\protect \) in
terms of \protect\( SU(3)\protect \) representations.}
\end{figure}

Now we consider the quantum operator \( \hat{H}_{1} \) obtained from
the matrix valued function \( H_{1} \) by replacing \( Z_{k},\overline{Z_{k}} \)
by operators \( a_{k},a_{k}^{+} \) respectively. We expect that \( \hat{H}_{1} \)
has two different eigenvalues close to \( E_{Line}=1 \), and \( E_{Orth}=0 \)
(in the semi-classical limit \( N\rightarrow \infty  \)). From its
invariance under the SU(3) group action, the eigenspaces of \( \hat{H}_{1} \)
are irreducible representations of this group with respective dimensions
\( {\mathcal{N}}_{Line},{\mathcal{N}}_{Orth} \). The eigenspaces
of \( \hat{H}_{1} \) are obtained by Young tableau technics, see
figure \ref{f:fig1}. The Weyl's formula gives \begin{eqnarray*}
{\mathcal{N}}_{Line}=\left( N+2\right) \left( N+3\right) /2, &  & \\
{\mathcal{N}}_{Orth}=N\left( N+2\right) . &  & 
\end{eqnarray*}

\begin{figure}[htbp]
{\centering \resizebox*{0.6\columnwidth}{!}{\includegraphics{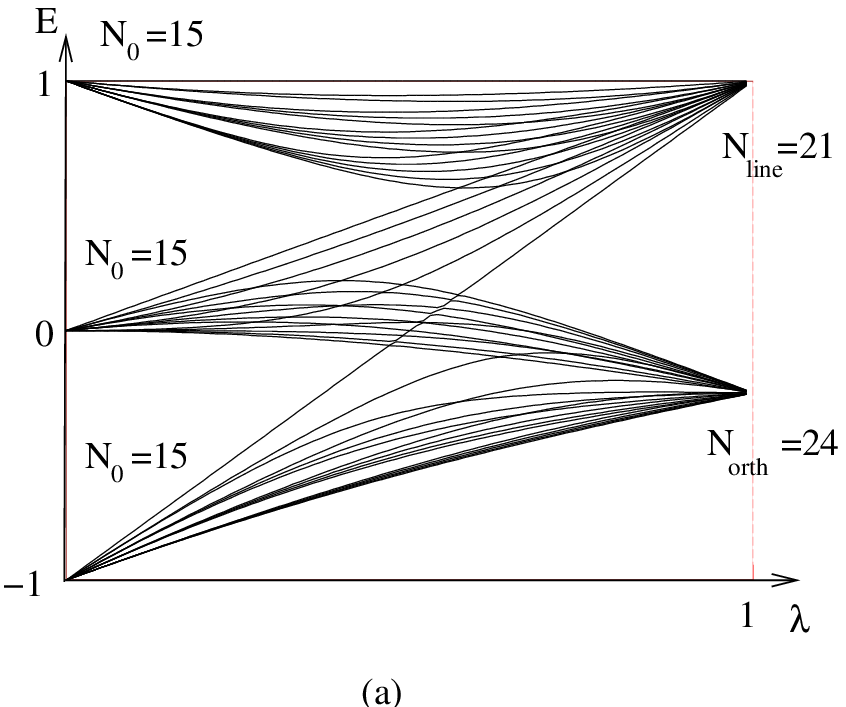}} \\
\resizebox*{0.6\columnwidth}{!}{\includegraphics{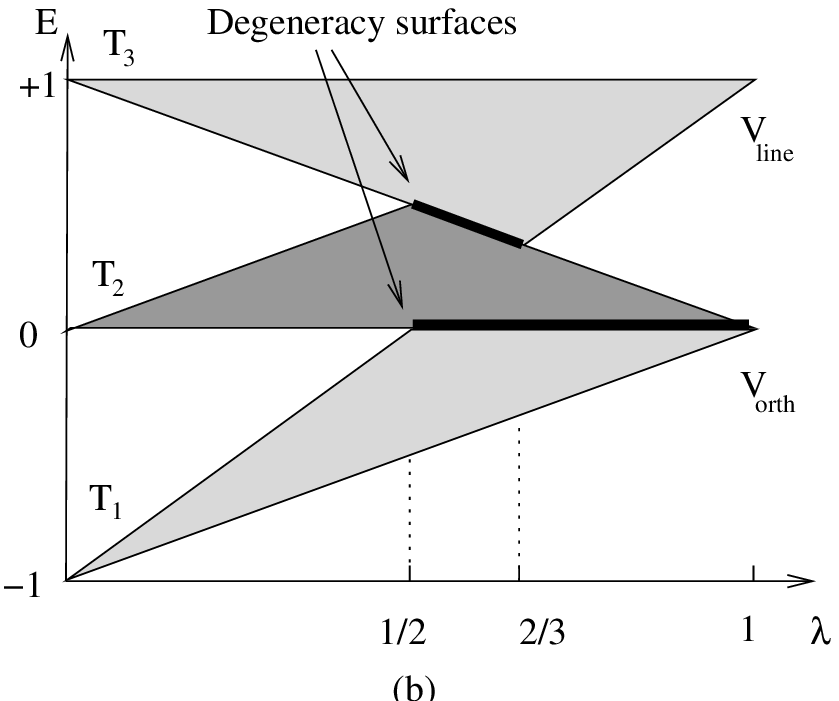}} \par}

\caption{\label{f:fig2}(a) The exact spectrum of \protect\( \hat{H}_{\lambda }\protect \),
for \protect\( N=4\protect \), with a transfer of \protect\( \Delta N=N+2=6\protect \)
levels towards the upper group. \protect \\
(b)The continuous bands spectrum in the Born-Oppenheimer approximation,
interpreted as vector bundles. The Atiyah-Singer formula (\ref{e:atiyah-singer})
gives the number of levels of a given band, say \protect\( N_{Line}\protect \),
in terms of the topology of the vector bundle \protect\( V_{Line}\protect \). }
\end{figure}

The splitting of the whole system of \( 3(N+1)(N+2)/2 \) energy levels
into two bands is the result of the SU(3) symmetry of the model Hamiltonian
\( \hat{H}_{1} \). We can now introduce the perturbation which breaks
the SU(3) symmetry, and analyses the persistence of the two bands
and the possible evolution of the band structure. In order to see
qualitative modifications of the band structure, we introduce an external
parameter \( \lambda  \) and construct a one-parameter family of
Hamiltonians \( H_{\lambda } \), \( \lambda \in \left[ 0,1\right]  \),
which relates the Hamiltonian \( H_{1} \) introduced above, with
a Hamiltonian \( H_{0} \) which does not depend on \( \left[ Z\right]  \)
(no vibronic coupling) but possess three non-degenerate electronic
states: \begin{eqnarray}
H_{\lambda }\left( \left[ Z\right] \right) =\left( 1-\lambda \right) H_{0}+\lambda H_{1}\left( \left[ Z\right] \right) , &  & \label{ParFam} \\
\textrm{with }H_{0}=Diag\left( -1,0,1\right) . &  & \nonumber 
\end{eqnarray}
 For \( \lambda =0 \), every eigenvalue \( E=-1,0,1 \) of the Hamiltonian
\( \hat{H}_{0} \) has multiplicity \( {\mathcal{N}}_{0} \). As \( \lambda  \)
varies from \( 0 \) to \( 1 \), we get a redistribution of the energy
levels from three groups of levels towards two groups, as shown on
figure \ref{f:fig2}(a).

 The physical interpretation of the family of Hamiltonians (\ref{ParFam})
can be done for both realizations mentioned above, the vibrational
structure of the combination bands and the Jahn-Teller effect.

 In the case of vibrational combination bands the \( \hat{H}_{0} \)
Hamiltonian corresponds to the vibrational structure formed by overtones
of triply degenerate harmonic oscillator coupled with three close
but non-degenerate vibrational states with higher frequencies. The
\( \hat{H}_{1} \) term introduces the interaction between low and
high frequency states. It becomes leading term as soon as the splitting
between high frequency states goes to zero or equivalently as soon
as the coupling between low and high frequency motion becomes more
important than the high frequency splitting. This means, for example,
that the formal parameter \( \lambda  \) in Eq. (\ref{ParFam}) can
be associated with the polyad quantum number \( N \) which is naturally
responsible for the strength of the interaction between slow and fast
subsystems.

For the Jahn-Teller \( F-f \) problem the Hamiltonian \( \hat{H}_{1} \)
appears in a natural way as a diagonal in polyad quantum number effective
operator. At the same time the \( \hat{H}_{0} \) limit formally corresponds
to the absence of the Jahn-Teller effect and even to the absence of
the three-fold degeneracy which is the origin of the Jahn-Teller interactions.
To give again a physical significance to the family of Hamiltonians
(\ref{ParFam}), one should introduce the reason of the splitting
of electronic states. This can be formally done by adding some external
magnetic field. In such a case, to approach the \( \hat{H}_{0} \)
limit, the splitting caused by magnetic field should become larger
than the splitting of vibrational polyads caused by pseudo Jahn-Teller
interactions, i.e. by vibronic interactions between different non-degenerate
electronic states. Similar effect can be produced by any symmetry
breaking effect whose influence is much more important on electronic
subsystem than on vibrations.

\subsection{ Symmetry analysis of band decompositions}

From the point of view of the Jahn-Teller dynamic effect the most
interesting prediction of our topological analysis is the splitting
of the whole multiplet of \( 3(N+1)(N+2)/2 \) vibronic states into
two groups which we will interpret later as an elementary band {}``Line''
and a topologically coupled band {}``Orth.'', both with non trivial
topological Chern class. From the point of view of practical applications
of this statement to concrete molecules, the simple number of states
is clearly an insufficient information in order to interpret the observed
energy level pattern formed by two different bands. A much more detailed
and useful description of two bands can be done by indicating the
symmetry of all levels forming each band. In the case of high symmetry
molecules this information significantly reduces possible misinterpretation
of the band structures in concrete examples. This is why, before going
to more general topological characterization, we look now at the problem
of the decomposition of the whole set of energy levels into different
bands, from the point of view of different symmetry groups. 

The highest symmetry is the SU(3) symmetry which is the dynamic symmetry
group of the 3-D isotropic harmonic oscillator (\ref{e:Hvib}). The
actual invariance group of the concrete molecular problem is a finite
point symmetry group \( G \) (tetrahedral \( G=T_{d} \) or octahedral
\( G=O_{h} \) for examples mentioned in this paper). The very convenient
approximate intermediate symmetry group O(3) is quite useful to give
additional label for groups of levels. The classification based on
the chain of group SU(3)\( \supset O(3)\supset G \) enables one to
verify the presence of groups of levels which can be interpreted as
bands.

In the case of \( F_{2}-f_{2} \) Jahn-Teller effect (the coupling
of triply degenerate electronic state of \( F_{2} \) symmetry with
vibrational polyads formed by triply degenerate vibrations of the
same symmetry for molecule with \( T_{d} \) symmetry group) the splitting
into two bands associated with two irreducible representations of
SU(3) dynamical symmetry group is followed by further decomposition
into irreducible representations of O(3) and \( T_{d} \) groups.
For low vibrational polyads this decomposition is given in Table \ref{T:splitt}
through the decomposition of each band into irreducible representations
of the symmetry group \( T_{d} \) and intermediate O(3) symmetry
group.

\begin{table}[htbp]

\caption{Splitting of multiplets of vibrational polyads into two bands for
weak \protect\( F_{2}-f_{2}\protect \) Jahn-Teller problem. Energy
levels forming each band are listed by their symmetry labels for polyads
with \protect\( N=1,2,3,4,5\protect \). First sub-line gives for
each band the decomposition into irreducible representations of \protect\( T_{d}\protect \)
symmetry group. Second sub-line gives similar decomposition in the
case of \protect\( O(3)\protect \) invariance group.}

\label{T:splitt}

\begin{tabular}{l|c|ccc}
  Band &
  Degen &
  \( N=1 \)&
  \( N=2 \)&
  \( N=3 \)\\
\hline
 Line &
  \( (N+2)(N+3)/2 \)&
  \( A_{1},E,F_{2} \)&
  \( A_{1},F_{1},2F_{2} \)&
  \( 2A_{1},2E,F_{1},2F_{2} \)\\
&
&
  \( 0_{g},2_{g} \)&
  \( 1_{u},3_{u} \)&
  \( 0_{g},2_{g},4g \)\\
\hline
 Orth. &
  \( N(N+2) \)&
  \( F_{1} \)&
  \( E,F_{1},F_{2} \)&
  \( A_{2},E,2F_{1},2F_{2} \)\\
&
&
  \( 1_{g} \)&
  \( 1_{u},2_{u} \)&
  \( 1_{g},2_{g},3_{g} \)\\
\end{tabular}

\( \qquad \qquad \qquad \qquad \qquad \qquad  \)\begin{tabular}{cc}
 \( N=4 \)&
 \( N=5 \)\\
\hline
  \( A_{1},E,2F_{1},4F_{2} \)&
  \( 2A_{1},A_{2},3E,2F_{1},4F_{2} \)\\
  \( 1_{u},3_{u},5_{u} \)&
 \( 0_{g},2_{g},4_{g},6_{g} \)\\
\hline
  \( A_{1},A_{2},2E,3F_{1},3F_{2} \)&
  \( A_{1},A_{2},3E,5F_{1},4F_{2} \)\\
  \( 1_{u},2_{u},3_{u},4_{u} \)&
  \( 1_{g},2_{g},3_{g},4_{g},5_{g} \)\\
\end{tabular}
\end{table}

The decomposition into two bands with \( (N+2)(N+3)/2 \) and \( N(N+2) \)
energy levels is imposed by the approximate dynamical SU(3) symmetry
which becomes an exact symmetry for the Hamiltonian (\ref{ParFam})
in the limit \( \lambda =1 \). The important point now is the justification
of the persistence of this decomposition found in the absence of SU(3)
symmetry, on the basis of topological arguments.

\section{Topological interpretation}

We now proceed to the interpretation of the redistribution phenomenon
observed in figure \ref{f:fig2}(a) within the Born-Oppenheimer approximation
using the semi-quantum model. The matrix \( H_{\lambda }\left( \left[ Z\right] \right)  \)
has three eigenvalues \( E_{1}\left( \lambda ,\left[ Z\right] \right) \leq E_{2}\left( \lambda ,\left[ Z\right] \right) \leq E_{3}\left( \lambda ,\left[ Z\right] \right)  \)
forming three bands when \( [Z]\in {\mathbb {C}}P^{2} \) varies,
see figure \ref{f:fig2}(b). The vibrational motion is now treated
classically, and so the fine structure of figure \ref{f:fig2}(a)
(due to quantization of the vibrations), has been replaced by a continuous
distribution of energies. Degeneracies between consecutive bands occur
for certain values of \( \left( \lambda ,\left[ Z\right] \right)  \).
Between bands 2 and 3, there is a degeneracy surface in the range
\( 1/2<\lambda <2/3 \) %
\footnote{ For a general family of Hermitian matrices, degeneracies occur with
three external parameters; this is the Wigner-Von Neumann theorem,
see \cite{key-6}, §79. Here, with the five external parameters \( \left( \lambda ,\left[ Z\right] \right)  \),
degeneracies are expected to hold on two-dimensional surfaces. The
degeneracy surface between bands 2 and 3 is a homologically non trivial
sphere \( S^{2} \) in \( {\mathbb {R}}\times {\mathbb {C}}P^{2} \). 
} . For a fixed value of \( \lambda  \), the associated eigenspaces
of the eigenvalues \( E_{1},E_{2},E_{3} \) also depend on \( \left[ Z\right]  \)
and therefore define vector bundles%
\footnote{A vector bundle \( F \) over a space \( P \) is a family of vector
space \( F(p) \) which depend continuously on a parameter \( p\in P \).
In the Born-Oppenheimer approximation, vector bundles appear naturally:
the vector space is spanned by the stationary state(s) of the fast
motion, which depends on the slow classical state \( p\in P \) in
phase space. The topology of a vector bundle reflects the twists among
the space \( F(p) \) as \( p \) is varied. A vector bundle with
no topology (no twist) is called trivial. 
}  over \( {\mathbb {C}}P^{2} \). The rank \( r \) of the bundle
is the dimension of these eigenspaces.

The topology of a general vector bundle \( F \) over \( {\mathbb {C}}P^{2} \)
is characterized by its rank \( r \) and a polynomial \( C(F)=1+Ax+Bx^{2} \),
called the total Chern class, with integer coefficients \( A,B\in {\mathbb {Z}} \),
see \cite{eguchi} p.300. A rank \( r=1 \) bundle has necessarily
\( B=0 \). The Chern class of the sum \( V\oplus V' \) of two vector
bundles is the product of polynomials \( C(V\oplus V')=C(V)\wedge C(V') \),
keeping only terms of degree not greater than 2.

For \( \lambda <1/2 \), on figure \ref{f:fig2}(b), there are three
distinct bands, and so three associated vector bundles of rank \( r=1 \),
noted \( T_{1},T_{2},T_{3} \). These bundles are trivial in the sense
that for \( \lambda =0 \) the eigenspaces do not depend on \( \left[ Z\right]  \),
and by continuity the topology is also trivial for any \( \lambda <1/2 \).
The Chern classes are \( C(T_{i})=1 \), with \( A=B=0 \).

 For \( 1/2<\lambda <2/3 \), there are degeneracy surfaces, and the
three previous bundles \( T_{1},T_{2},T_{3} \) {}``interact'' together
and form a vector bundle \( T=T_{1}\oplus T_{2}\oplus T_{3} \) of
rank 3. This is exactly the vector space \( {\mathcal{H}}_{elec}={\mathbb {C}}^{3} \)
which does not depend on \( \left[ Z\right]  \), so \( T \) is also
trivial, this confirms that \( C(T)=1 \).

 For \( 2/3<\lambda <1 \), the previous bundle \( T \) decomposes
into two bundles \( T=V_{Line}\oplus V_{Orth} \). The upper energy
band corresponds to a rank 1 vector bundle \( V_{Line} \). By continuity,
its topology can be computed at \( \lambda =1 \), and corresponds
to the well known non trivial canonical bundle over \( {\mathbb {C}}P^{2} \)
with \( C(V_{Line})=1+x \), see \cite{eguchi}. The lower band corresponds
to the orthogonal bundle \( V_{Orth} \). This is a non trivial bundle
of rank \( r=2 \), with \( C(V_{Orth})=1-x+x^{2} \), because \( 1=C\left( T\right) =C(V_{Line})\wedge C(V_{Orth}) \).
The vector bundle \( V_{Orth} \) cannot be decomposed into two rank
1 bundles just because its Chern class can not be factorized, \( 1-x+x^{2}\neq (1+Ax)\wedge \left( 1+A'x\right)  \)
with integers \( A,A' \). Therefore we obtain that \( V_{Orth} \)
is composed of \emph{two bands which are topologically coupled}: a
continuous perturbation cannot create a gap in it. To be decoupled,
they would need a topological interaction with an external third band.
This interesting phenomenon seems to be new in molecular physics.

\subsection{ Number of energy levels in band and index theorem}

The relation between the topology of a vector bundle \( F \) defined
in the Born-Oppenheimer approximation, and the number of levels in
the exact energy spectrum is given by the remarkable index formula
of Atiyah-Singer (\cite{eguchi} p.330), which, in the case of the
molecular model studied in this paper, reads as follows: \begin{equation}
\label{e:atiyah-singer}
{\mathcal{N}}(F)=\left[ Ch\left( F\right) \wedge Ch\left( Polyad\right) \wedge Todd\left( {\mathbb {C}}P^{2}\right) \right] _{/x^{2}},
\end{equation}
 where \( Ch(F)=r+Ax+\frac{1}{2}\left( A^{2}-2B\right) x^{2} \) is
the Chern character for the topology of the vector bundle \( F \).
\( Ch\left( Polyad\right) =\exp \left( Nx\right)  \) characterizes
the quantization of the polyad space%
\footnote{ To be precise, \( Ch\left( Polyad\right) =\exp \left( Nx\right)  \)
is the Chern Character of a holomorphic line bundle which enters in
the geometric quantization description of the space \( {\mathcal{H}}_{polyad} \).
See \cite{key-7} chap.8. 
} , and \( Todd\left( {\mathbb {C}}P^{2}\right) =1+\frac{3}{2}x+x^{2} \)
is related to the topology of the phase space \( {\mathbb {C}}P^{2} \).
To make calculation using the right hand side of the formula (\ref{e:atiyah-singer}),
one has to multiply the three polynomials, and keep the coefficient
of the term \( x^{2} \). This gives

\ \[
{\mathcal{N}}(F)=\frac{r}{2}N^{2}+\left[ \frac{3r}{2}+A\right] N+\left[ r+\frac{3A}{2}+\frac{A^{2}}{2}-B\right] .\]
 It is easy to check that this formula gives the values of \( {\mathcal{N}}_{Line} \)
and \( {\mathcal{N}}_{Orth} \) computed before. The Young tableau
machinery was possible at \( \lambda =1 \) thanks to a special symmetry.
Here the index formula with its topological nature is much more general
and is valid for any values of \( \lambda  \). The topology modification
is related to the level exchange: in the interval \( 1/2<\lambda <2/3 \)
there is a transfer of \( \Delta N={\mathcal{N}}_{Line}-{\mathcal{N}}_{0}=N+2 \)
levels to the upper group, through the degeneracy surface%
\footnote{The value \( \Delta N=\left( N+1\right) +1 \) can be interpreted
as \( \left( N+1\right)  \) quantized states of the degeneracy sphere
plus an extra level due to the topology of the degeneracy bundle. 
}.

It is useful to note here that the number of quantum states for any
band defined in the classical limit over \( \mathbb {C}P^{2} \) is
a quadratic function of the polyad quantum number \( N \). The coefficients
of this quadratic polynomial give information about the total Chern
class and therefore the topology of the fiber bundle. In the completely
classical model, when both slow and fast dynamical variables are treated
through classical mechanics, a similar relation between the volume
of the reduced classical phase space and Chern classes was established
by Duistermaat and Heckman \cite{DH}.

In comparison with the ro-vibrational model described on a \( S^{2} \)
phase space of dimension \( 2 \) \cite{fred-boris}, this model is
much more rich due to the dimension \( 4 \) of the phase space \( {\mathbb {C}}P^{2} \).
The new phenomenon consists in a formation of the degeneracy surface
together with the transition of a whole group of levels, instead of
formation of isolated degeneracy points with transition of one level
through each of them, for the case of \( S^{2} \) phase space. The
band topology is described by two integers (first and second Chern
classes) which play the role of the topological quantum numbers for
energy bands. Topologically coupled energy bands (i.e. non decomposable
vector bundles), appear for problems with four-dimensional \( {\mathbb {C}}P^{2} \)
phase space, whereas for \( S^{2} \) phase space one needs only a
single Chern index to characterize the coupling with fast quantum
subsystem, and consequently every elementary band has rank 1.

 It should be noted that the coupling of a slow classical subsystem
with at least three electronic quantum states is a necessary condition
to see the formation of topologically non trivial bands from initially
isolated trivial bands. In the above model, two isolated trivial bands
can not be transformed into non-trivial bands for topological reasons
\cite{AAM}. This is just because the Chern class of a trivial rank
2 bundle over \( \mathbb {C}P^{2} \) is \( C(T)=1 \), and the decomposition
into two rank one bundles \( T=F_{1}\oplus F_{2} \) gives the factorization
\( C(\mathbb {C}^{2})=1=(1+A_{1}x)(1+A_{2}x) \) which implies that
\( A_{1}=A_{2}=0 \), and that \( F_{1} \) and \( F_{2} \) are trivial
bundles. This constraint does not exist in the case of vector bundles
over sphere \( S_{2} \), where the Chern class is just a degree one
polynomials (one just has then \( A_{1}=-A_{2} \)).

This precedent remark does not mean that two isolated trivial bands
cannot become coupled due to vibronic interaction: the formation of
degeneracy surfaces between two bands is naturally possible.

\section{ Conclusions}

 The model presented in this paper can be adapted to different situations
in molecular physics where the slow motion phase space can be \( S^{2},{\mathbb {C}}P^{2},{\mathbb {C}}P^{3},\ldots  \)
and products of them, depending of the relevant degrees of freedom,
which can be rotational, vibrational, or electronic. As a concrete
molecular object which fits well the assumptions made by the present
model, we can mention the ro-vibronic structure of Jahn-Teller molecule
V(CO)\( _{6} \) \cite{key-4}. This concrete example needs a slight
extension to the model with a \( {\mathbb {C}}P^{2}\times S^{2} \)
slow phase space coupled with three quantum states, and is especially
interesting for further analysis.

 Our approach can also be used in other areas of physics where Born-Oppenheimer
approximation is relevant and where the slow subsystem can be considered
as classical and coupled to several quantum states.

>From the mathematical point of view we have shown a concrete manifestation
of Atiyah-Singer's index formula in molecular physics. This gives
new field of applications to the index formula. Up to now it was mainly
related to quantum chromodynamics with the topology of instantons
and gluon fields, but for different reasons (\cite{eguchi} p.355).
The topology of bands are also known to have an importance in solid
state physics for the integer quantum Hall effect \cite{thouless2,fred3}.
Molecular systems supply a lot of new examples where the qualitative
and quantitative predictions of Atiyah-Singer's index formula can
be unambiguously tested and verified with high accuracy experimental
measurements and numerical calculations.

\paragraph{Acknowledgements}

 The work is supported by the EU project Mechanics and Symmetry in
Europe contract \#HPRN-CT-2000-00113.

\end{document}